\documentclass[aps,prl,preprint,groupedaddress]{revtex4}
\usepackage{graphicx}
\begin{document}


\title{ Bijectivity of the Normalization and Fermi-Coulomb Hole Sum Rules for Approximate Wave Functions

 }


\author{Xiao-Yin Pan}
\author{ Viraht Sahni}
\affiliation{ The Graduate School of the City University of New
York, New York, New York 10016. }
\author{Lou Massa}
\affiliation{ The Graduate School of the City University of New
York, New York, New York 10016. }

\date{\today}

\begin{abstract}
We prove the bijectivity of the constraints  of  normalization
and  of the Fermi-Coulomb hole charge sum rule at each electron
position for approximate wave functions.  This bijectivity is
surprising in light of the fact that normalization depends upon
the probability of finding an electron at
  some position, whereas the Fermi-Coulomb hole sum rule depends on the
  probability of two electrons staying apart because of correlations due to the
  Pauli exclusion principle and Coulomb repulsion.
  We further demonstrate the bijectivity of these sum rules by example.

\end{abstract}

\pacs{}

\maketitle


Sum rules play an important role in physics, and there are many ways
in which they are employed.   Within the realm of electronic structure
 theory, for example, accurate properties of a system may be determined
 by the use of an approximate potential whose parameters are adjusted
 so as to ensure the satisfaction of a sum rule.  Metal surface
  properties such as the surface energy and work function are obtained
   by application of the Theophilou-Budd-Vannimenus sum rule \cite{1} which
    relates the value of the electrostatic potential at the surface to
    the known bulk properties of the metal.  The parameters in a model
    effective potential at a metal surface are then adjusted \cite{2} so
     as to satisfy this sum rule.  Another manner in which sum rules have
      proved to be significant is in the context of Kohn-Sham density functional
       theory (KS-DFT) \cite{3}, a local effective potential theory of
       electronic structure that is extensively employed in atomic, molecular,
        and condensed matter physics.  In KS-DFT, all the many-body effects are
        incorporated in the `exchange-correlation' energy functional of the
        ground state density.  Since this functional is unknown, it must be approximated.
        A successful approach \cite{4} to the construction of approximate `exchange-correlation'
        energy functionals, and of their derivatives which represent the local effective potential
         in the theory, is the requirement of satisfaction of various scaling laws \cite{5} together
          with those of sum rules on the Fermi and Coulomb hole charge distributions \cite{6}.
           In the recently developed Quantal density functional theory (Q-DFT) \cite{6}, the local
           effective potential is described instead in terms of the system wave function.
           Thus, one method for the construction of the local effective potential in Q-DFT is to
           employ an approximate wave function that is a functional of some functions \cite{7}.
           These latter functions are determined such that the wave function functional satisfies
           various constraints such as normalization, the Fermi-Coulomb or Coulomb hole sum rules,
           or reproduces a physical observable of interest such as the density, diamagnetic susceptibility,
           nuclear magnetic constant, etc.\cite{7}.  \\

The satisfaction of a particular sum rule by an approximate
potential, or an `exchange-correlation' energy functional, or a
wave function functional, however, does not necessarily imply the
satisfaction of other sum rules.   In this paper we describe a
counter intuitive \emph{bijective} relationship between the sum
rules of normalization and that of the Fermi-Coulomb or Coulomb
hole charge.  The satisfaction of either one of the sum rules by
an \emph{approximate wave function} ensures the satisfaction of
the other. This bijectivity is counter intuitive because the
constraints of normalization and of the Fermi-Coulomb hole depend
on distinctly different quantum-mechanical probabilities. The
bijectivity is also of importance from a practical numerical
perspective.  The proof and demonstration of the bijectivity of
these sum rules constitutes the paper.\\

To understand why this bijectivity is so counter to intuition, let
us consider the physics underlying the two properties of an
electronic system that these sum rules depend upon.  For a system
of $N$ electrons, the constraint of normalization on an
approximate wave function $\Psi({\bf X})$ requires that \\
\begin{equation}
\int \Psi({\bf X})^{*} \Psi({\bf X})d{\bf X}=1,
\end{equation}
where ${\bf X} = {\bf x}_{1},...,{\bf x}_{N}; d {\bf X} = d {\bf
x}_{1},...,d{\bf x}_{N}; {\bf x} = {\bf r}, s$ with ${\bf r}$ and
$s$ being the spatial and spin coordinates of an electron. (Atomic
units $e = \hbar = m = 1$ are assumed.) Equivalently, this sum
rule may be written in terms of the electronic density $\rho({\bf
r})$. The density $\rho({\bf r})$ is $N$ times the probability of
an electron being at ${\bf r}$ :
\begin{equation}
\rho({\bf r})=N\sum_{i}\int \Psi^{*}({\bf r} \sigma,{\bf
X}^{N-1})\Psi({\bf r} \sigma,{\bf X}^{N-1}) d{\bf X}^{N-1} ,
\end{equation}
where $d {\bf X}^{N-1} = d{\bf x}_{2},...,d{\bf x}_{N}$.  The
normalization sum rule then becomes

\begin{equation}
\int \rho({\bf r})d{\bf r}=N .
\end{equation}

The density $\rho({\bf r})$ is a \emph{static} or \emph{local}
charge distribution.  By this is meant that its structure remains
unchanged as a function of electron position ${\bf r}$.
Integration of this charge distribution---the normalization sum
rule---then gives the number $N$ of electrons. Thus, normalization
is a statement as to the number of electrons in the system.\\

The definition of the Fermi-Coulomb hole charge distribution
$\rho_{xc}({\bf r} {\bf r}')$ derives from that of the
pair-correlation density $g({\bf r} {\bf r}')$. The
pair-correlation density is the density at ${\bf r}'$ for an
electron at ${\bf r}$. The density at ${\bf r}'$ differs from that
at ${\bf r}$ because of electron correlations due to the Pauli
exclusion principle and Coulomb repulsion.  Thus, the pair density
is defined as

\begin{equation}
g({\bf r} {\bf r}')=\langle \Psi| \sum_{i\neq j}\delta({\bf
r}_{i}-{\bf r}) \delta({\bf r}_{j}-{\bf r}')|\Psi\rangle/\rho({\bf
r}).
\end{equation}
Its total charge, for each electron position ${\bf r}$, is
therefore
\begin{equation}
\int g({\bf r} {\bf r}') d{\bf r}'=N-1.
\end{equation}

The pair-correlation density $g({\bf r} {\bf r}')$ is a
\emph{dynamic} or \emph{nonlocal} charge distribution in that its
structure changes as a function of electron position for
\emph{nonuniform} electron density systems.  If there were no
electron correlations, the density at ${\bf r}'$ would be
$\rho({\bf r}')$. Hence, the pair-correlation density is the
density at ${\bf r}'$ plus the reduction in density at ${\bf r}'$
due to the electron correlations.  The reduction in density about
an electron which occurs as a result of the Pauli exclusion
principle and Coulomb repulsion is the Fermi-Coulomb hole charge
$\rho_{xc}({\bf r} {\bf r}')$. Thus, the Fermi-Coulomb hole is
defined as
\begin{equation}
\rho_{xc}({\bf r} {\bf r}')= g({\bf r} {\bf r}')-\rho({\bf r}').
\end{equation}
The Fermi-Coulomb hole $\rho_{xc}({\bf r} {\bf r}')$ about an
electron is also a \emph{dynamic} or \emph{nonlocal} charge
distribution. For nonuniform electron gas systems, its structure
is different for each electron position. Since each electron digs
a hole in the inhomogeneous sea of electrons equal in charge to
that of a proton, it follows that the total charge of the
Fermi-Coulomb hole surrounding an electron, \emph{for each
electron position} ${\bf r}$, is
\begin{equation}
\int \rho_{xc}({\bf r} {\bf r}') d{\bf r}'=-1.
\end{equation}
This is the Fermi-Coulomb hole sum rule.
\\

The definition of the Coulomb hole $\rho_{c}({\bf r} {\bf r}')$,
which is the reduction in density at ${\bf r'}$ for an electron at
${\bf r'}$ because of Coulomb repulsion, in turn derives from that
of the Fermi-Coulomb $\rho_{xc}({\bf r} {\bf r}')$ and   Fermi
$\rho_{x}({\bf r} {\bf r}')$ holes.  The Fermi hole is the
reduction in density at ${\bf r'}$ for an electron at ${\bf r}$
that occurs due to the Pauli exclusion principle.   The Fermi hole
is defined via the pair-correlation density $g_{s}({\bf r} {\bf
r}')$ derived through a normalized Slater determinant
$\Phi\{\varphi_{i} \}$ of single particle orbitals
$\varphi_{i}({\bf x})$:

\begin{eqnarray}
g_{s}({\bf r} {\bf r}')&=&\frac{\langle \Phi\{\varphi_{i} \}|
\sum_{i\neq j}\delta({\bf r}_{i}-{\bf r}) \delta({\bf r}_{j}-{\bf
r}')|\Phi\{\varphi_{i} \}\rangle} {\rho({\bf r})}\\
 &=&\rho({\bf
r}')+\rho_{x}({\bf r}{\bf r}').
\end{eqnarray}
The orbitals $\varphi_{i}({\bf x})$ may be generated either
through KS-DFT or Q-DFT in which case the density $\rho({\bf r})$
is the same as that of the interacting system, or they could be
the Hartree-Fock theory orbitals for which the density is
different. As the sum rule on $g_{s}({\bf r} {\bf r}')$ is the
same as in Eq. (5), and the Slater determinant is normalized, the
total charge of the Fermi hole, for \emph{each electron position
${\bf r}$}, is also that of a proton:
\begin{equation}
\int \rho_{x}({\bf r}, {\bf r}') d{\bf r}'=-1.
\end{equation}
The Coulomb hole is then defined as the difference between the
Fermi-Coulomb and Fermi holes:
\begin{equation}
 \rho_{c}({\bf r} {\bf r}') =
 \rho_{xc}({\bf r} {\bf r}') - \rho_{x}({\bf r} {\bf r}').
\end{equation}.
The total charge of the Coulomb hole, \emph{for each electron
position ${\bf r}$}, is therefore zero:
\begin{equation}
\int \rho_{c}({\bf r} {\bf r}') d{\bf r}'=0.
\end{equation}
This is the Coulomb hole sum rule.
\\

Both the normalization and the Fermi-Coulomb or Coulomb hole
constraints are charge conservation sum rules. However, their
physical origin, and therefore the charge conserved in each case,
is different.  That these distinctly different charge conservation rules are intrinsically linked bijectively constitutes
 the theorem we prove.\\

\emph{ Theorem}: The normalization and Fermi-Coulomb or Coulomb
hole sum rules are bijective. Satisfaction of the normalization
sum rule by an \emph{approximate wave function} implies the
automatic satisfaction of the Fermi-Coulomb or Coulomb hole sum
rules \emph{for each electron position}.  Conversely, the
satisfaction of the Fermi-Coulomb or Coulomb hole sum rules
\emph{for each electron position} by an \emph{approximate wave
function} implies the normalization of that wave function:
\begin{equation}
 \left ( \begin{array}{c} \int \Psi({\bf X})^{*} \Psi({\bf X})d{\bf X}=1\\ or \\ \int \rho({\bf r})d{\bf r}=N  \end{array}  \right )
 \begin{array}{c}\rightarrow \\\leftarrow \end{array} \left ( \begin{array}{c} \int \rho_{xc}({\bf r} {\bf r}') d{\bf
r}'=-1 \\ or\\\int \rho_{c}({\bf r} {\bf r}') d{\bf r}'=0
\end{array}\right )
\end{equation}
\\

\emph{Proof}: (a)The proof of the arrow to the right in Eq. (13)
is as follows. Let us assume an approximate wave function that is
normalized. Then,  integration of Eq.(6) over ${\bf r}'$ using the
normalization constraint of Eq.(3) leads directly to the
Fermi-Coulomb hole sum rule of Eq.(7).\\
(b) For the arrow to the left, consider an approximate wave
function that satisfies the Fermi-Coulomb hole sum rule Eq.(7) for
each electron position ${\bf r}$. The sum rule Eq.(5) on the
pair-correlation density $g({\bf r} {\bf r}')$ follows from its
definition Eq.(4) \emph{which is independent of whether or not the
wave function is normalized}.  Thus, since both the sum rules on
the Fermi-Coulomb hole and the pair-correlation density are
satisfied, then on integration of  Eq.(6) over ${\bf r}'$,
normalization of the wave function is ensured. \\
(c) Consider an approximate wave function from which one
constructs a Fermi-Coulomb hole for each electron position ${\bf
r}$. For a normalizd  Slater determinant $\Phi\{\varphi_{i} \}$,
next define a Fermi hole $\rho_{x}({\bf r} {\bf r}')$ which then
satisfies the Fermi hole sum rule of Eq.(10). If the satisfaction
of the Coulomb hole sum rule is now ensured, then this guarantees
the satisfaction of the Fermi-Coulomb hole sum rule, which as
shown in (b), ensures that the wave function is normalized.\\

Recall that normalization depends upon the probability of finding
an electron at some position.  On the other hand, the
Fermi-Coulomb and Coulomb hole sum rules depend on the reduction
in probability of two electrons approaching  each other.  The fact
that satisfaction of the integral condition of either one of these
probabilities means the satisfaction on the integral condition of
the other is not obvious, and therefore surprising.
\\

We next demonstrate the bijectivity of Eq. (13) by application to
the ground state of the Helium atom. The nonrelativistic
Hamiltonian of the atom is
\begin{equation}
\hat{H}=-\frac{1}{2}\nabla_{1}^{2}-\frac{1}{2}\nabla_{2}^{2}-\frac{Z}{r_{1}}
  -\frac{Z}{r_{2}}+\frac{1}{r_{12}},
\end{equation}
where ${\bf r}_{1}$, ${\bf r}_{2}$ are the coordinates of the two
electrons, $r_{12}$ is the distance between them, and $Z = 2$ is
the atomic number. The equivalence from left to right of Eq. (13)
can be easily demonstrated by assuming an approximate wave
function $\psi$ with parameters $c_{i} (i = 1,...,p)$ that is
normalized in the standard manner at the energy minimized values
of the parameters: $\frac{\partial I[\psi]}{\partial c_{i}} = 0$,
where $I[\psi] = \int\psi^{*}H \psi d\tau/ \int \psi^{*} \psi
d\tau $. On the other hand, the equivalence from right to left is
not as readily accomplished through such a wave function since the
Fermi-Coulomb hole sum rule must be satisfied at \emph{each}
electron position.  It is, however, possible to demonstrate the
bijectivity by assuming the wave function to be a functional of a
set of functions $\chi$: $\psi = \psi[\chi]$ instead of simply a
function. The functions $\chi$ are determined so as to satisfy the
normalization or
Fermi-Coulomb hole sum rules as described in Ref.7.\\

\begin{table}
\caption{ The satisfaction of the Coulomb hole sum rule Eq.(12)
for different electron positions ${\bf r}$ \cite{9}.}
\renewcommand{\arraystretch}{0.4}
\begin{tabular}{|c|c|}\hline
 ${\bf r}$(a.u.)& $\;\;\;\;\; \int \rho_{c}({\bf r}{\bf
r}^{\prime})d{\bf r}^{\prime}\;\;\;\;$  \\ \hline\hline

0.00566798  &  -0.00039251   \\\hline 0.13567807  &   0.00032610 \\
\hline 0.57016010  &   0.00034060   \\ \hline 0.72285115  &
0.00013025   \\\hline 0.89208965  &   0.00001584   \\\hline
1.07722084  &   0.00007529   \\\hline 1.49223766  &   0.00029097
\\\hline
1.96148536  &   0.00034743   \\\hline 3.91996382  &   0.00032567
\\\hline 5.15549169  &   0.00057862   \\ \hline
\end{tabular}
\end{table}
For the left to right equivalence, we choose the wave function
functional to be of the form \cite{7}
\begin{equation}
\psi[\chi]=\Phi(\alpha, s)[1-f(\chi; s,u)],
\end{equation}
with$ \Phi[\alpha,s]=(\alpha^{3}/\pi)e^{-\alpha s}$, $f(s,u)=e^{-q
u}(1+qu)[1-\chi(q;s,u)(1+u/2)]$, where $\alpha$ and $q$ are
variational parameters, $s=r_{1}+r_{2}, u = r_{12}$. The function
$\chi =\chi_{2}$ of \cite{7}, with the energy minimized values of
the parameters being $\alpha = 1.6629, q = 0.17049$. This wave
function is normalized to unity, the function $\chi$ being
determined as a solution to a quadratic equation. We further
assume, as in local effective potential energy theory, that the
Fermi hole $\rho_{x}({\bf r} {\bf r'}) = - \rho( {\bf r'})/2$. The
corresponding Coulomb holes $\rho_{c}({\bf r} {\bf r'})$ are
plotted in Figs. $1$ and $2$ for electron positions at ${ r} = 0,
0.566, 0.8, 1.0$ (a.u.) together with the exact Coulomb hole
\cite{8}. (The electron is on the $z$ axis corresponding to $
\theta= 0$. The cross section through the Coulomb hole plotted
corresponds to $\theta' = 0$ with respect to the electron-nucleus
direction.  The graph for $ r' < 0$ corresponds to the structure
for $\theta' = \pi$ and ${ r}'> 0$.) The two Coulomb holes, though
similar are inequivalent. Integration of both the exact and
approximate Coulomb holes for each electron position leads to a
total charge
of zero.\\

\begin{figure}
 \begin{center}
 \includegraphics[bb=0 0 570 772, angle=0, scale=0.4]{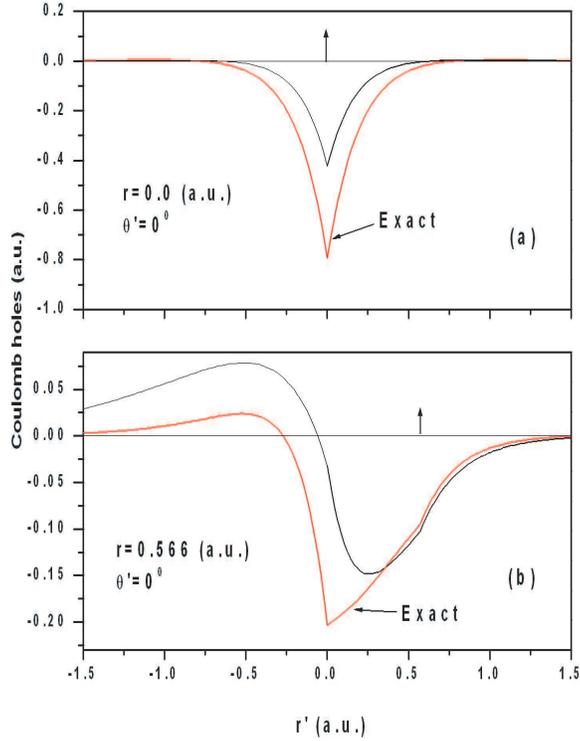}
 \caption{Cross-section through the Coulomb holes for electron positions at (a)$r=0$ (a.u.), and (b) $r=0.566$ (a.u.).
  The holes determined by the wave function functional of Eq.(15) and the `exact' hole are plotted . \label{}}
 \end{center}
 \end{figure}

 \begin{figure}
 \begin{center}
 \includegraphics[bb=0 0 580 775, angle=0, scale=0.4]{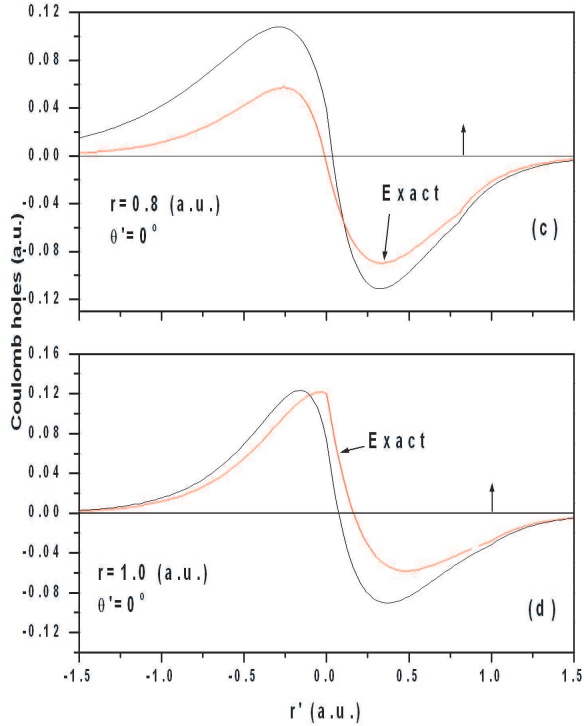}
 \caption{The figure caption is the same as in Fig.1 except that the cross sections plotted are
 for electron positions at (c)$r=0.8$ (a.u.), and (d) $r=1.0$ (a.u.).
  . \label{}}
 \end{center}
 \end{figure}
For the right to left equivalence of Eq. (13), we choose a
different wave function functional \cite{9}:
\begin{equation}
\psi[\chi]=\Phi(\phi_{i})[1-f({\bf r}_{1}{\bf r}_{2} )],
\end{equation}
with $ f({\bf r}_{1}{\bf r}_{2})=e^{-\beta ^{2} r^{2}}
[1-\chi({\bf R})(1+r/2))]$ ,  ${\bf r} = {\bf r}_{1} - {\bf
r}_{2}$, ${\bf R} = ({\bf r}_{1} + {\bf r}_{2})/2$, $\beta=q
[\rho({\bf R})]^{1/3}$, $q$ a variational parameter, and
$\Phi(\phi_{i})$ the Hartree-Fock theory prefactor \cite{10}. The
satisfaction of the Coulomb hole sum rule requires the solution of
a nonlinear integral Fredholm equation of the first kind for the
determination of the function $\chi({\bf R})$. We have solved
\cite{9} the linearized version of this integral equation for
$\chi({\bf R})$. The satisfaction of the Coulomb hole sum rule for
typical electron positions for $q = 1$ is given in Table I.  (We
do not plot the corresponding Coulomb holes as they are very
similar to those of Figs. 1 and 2.)  The wave function functional
of Eq. (16) thus determined satisfies the normalization constraint
to the same degree of accuracy as that of the sum rule given in
Table I. Hence, the bijectivity of the normalization and Coulomb
hole sum
rules is demonstrated by example.\\

In conclusion, we have proved the bijectivity of the normalization
and Fermi-Coulomb or Coulomb hole sum rules for approximate wave
functions.  The bijectivity is also significant from a numerical
perspective because it is much easier to normalize a wave function
than to ensure the satisfaction of the Fermi-Coulomb or Coulomb
hole sum rules for each electron position.  As shown by the
examples, the determination of a wave function functional via
normalization requires the solution of a quadratic equation,
whereas that determined via satisfaction of the Coulomb hole sum
rule requires the solution of an integral equation.  On the other
hand we note that the wave function functionals, as determined by
satisfaction of the different sum rules, are different.  Hence,
the Fermi-Coulomb and Coulomb holes, and therefore how the
electrons are correlated, will be different depending upon which
sum rule is satisfied.  It is unclear as to whether a better
representation of the electron correlations is achieved by
satisfaction of the normalization sum rule or that of the
Fermi-Coulomb hole.  Finally, the bijectivity explains the results
of our analysis \cite{11} of the Colle-Salvetti wave function
functional \cite{12}.  This wave function, which constitutes the
basis for the most extensively used 'correlation' energy
functional in the literature, is of the same form as that of Eq.
(16) except that  $\chi({\bf R}) = \sqrt{\pi} \beta/ (1 +
\sqrt{\pi} \beta)$, $\beta=q [\rho^{HF}({\bf R})]^{1/3}$. In
analyzing this wave function we had noted that it was neither
normalized nor did it satisfy the Coulomb hole sum rule.  These
facts are consistent with the bijectivity theorem proved above.
The lack of satisfaction of either one of the constraints ensures
the lack of satisfaction of
the other. \\

\begin{acknowledgments}
This work was supported in part by the Research Foundation of
 CUNY. L. M. was supported in part by NSF through CREST, and by
 a ``Research Centers in Minority Institutions'' award, RR-03037,
 from the National Center for Research Resources, National
 Institutes of Health.
\end{acknowledgments}

\end{document}